\documentclass[pdflatex,sn-mathphys-num]{sn-jnl}

\usepackage{graphicx}%
\usepackage{multirow}%
\usepackage{amsmath,amssymb,amsfonts}%
\usepackage{amsthm}%
\usepackage{mathrsfs}%
\usepackage[title]{appendix}%
\usepackage{xcolor}%
\usepackage{textcomp}%
\usepackage{manyfoot}%
\usepackage{booktabs}%
\usepackage{algorithm}%
\usepackage{algorithmicx}%
\usepackage{algpseudocode}%
\usepackage{listings}%
\usepackage{lineno}

\newcommand{\add}[1]{\textcolor{black}{#1}}

\theoremstyle{thmstyleone}%
\theoremstyle{thmstyletwo}%

\theoremstyle{thmstylethree}%

\begin{document}

\title[Article Title]{How optimal control of polar sea-ice depends on its tipping points}


\author*[1]{\fnm{Parvathi} \sur{Kooloth}}\email{parvathi.kooloth@pnnl.gov}

\author*[1]{\fnm{Jian} \sur{Lu}}\email{jian.lu@pnnl.gov}

\author[1]{\fnm{Craig} \sur{Bakker}}\email{craig.bakker@pnnl.gov}

\author[2]{\fnm{Derek} \sur{DeSantis}}\email{ddesantis@lanl.gov}

\author[1]{\fnm{Adam} \sur{Rupe}}\email{adam.rupe@pnnl.gov}

\affil[1]{ \orgname{Pacific Northwest National Laboratory}, \orgaddress{\city{Richland}, \postcode{99354}, \state{WA}, \country{USA}}}

\affil[2]{\orgname{Los Alamos National Laboratory}, \orgaddress{\city{Los Alamos}, \postcode{87545}, \state{NM}, \country{USA}}}


\abstract{Several Earth system components are at a high risk of undergoing rapid and irreversible qualitative changes or `tipping', due to increasing climate warming. Potential tipping elements include Arctic sea-ice, Atlantic meridional overturning circulation, and tropical coral reefs. Amidst such immediate concerns, it has become necessary to investigate the feasibility of arresting or even reversing the crossing of tipping thresholds using feedback control. In this paper, we study the control of an idealized diffusive energy balance model (EBM) for the Earth’s climate; this model has two tipping points due to strong co-albedo feedback. One of these tipping points is a `small icecap' instability responsible for a rapid transition to an ice-free climate state under increasing greenhouse gas (GHG) forcing. We develop an optimal control strategy for the EBM under different climate forcing scenarios with the goal of reversing sea ice loss while minimizing costs. We find that effective control is achievable for such a system, but the cost of reversing sea-ice loss nearly quadruples
for an initial state that \add{has just tipped} as compared to a state before reaching the tipping point. \add{We also show that thermal inertia may delay tipping leading to an overshoot of the critical GHG forcing threshold. This may offer a short intervention window (overshoot window) during which the control required to reverse sea-ice loss only scales linearly with intervention time. While systems with larger system inertia may have longer overshoot windows, this increased elbow room comes with a steeper rise in the requisite control once the intervention is delayed past this window.} Additionally, we find that the requisite control to restore sea-ice is localized in the polar region. }




\maketitle

\section{Introduction}\label{sec1}

The year 2023 marked the warmest year on record yet, with the global mean temperature about 1.5$^\circ$C warmer than the pre-industrial level. Coincidentally, 1.5$^\circ$C is also the threshold proposed by the Paris Agreement \cite{change2018global}, beyond which there is a risk of triggering multiple tipping points in the climate system \cite{lenton2008tipping,armstrong2022exceeding,wang2023mechanisms}; near these points, small perturbations in the system can lead to substantial and irreversible changes. Examples of these `tipping elements' include Atlantic meridional overturning circulation (AMOC) \cite{ditlevsen2023warning}, West Antarctic ice sheet collapse \cite{feldmann2015collapse}, Arctic winter sea ice loss \cite{hezel2014modeled, hankel2021role}, Antarctic sea-ice loss \cite{hobbs2024observational} and tropical coral reef die-offs \cite{hughes2017coral}. The research community has called for urgent actions to avert the tipping point risks as the current climate policies, even if implemented successfully, are unlikely to limit the warning to 2$^\circ$C \cite{meinshausen2022realization}. 

The imminent risk of catastrophic climate transitions due to global warming has increased the urgency of exploring possible climate interventions, such as geoengineering, to stabilize the climate. Extensive exploratory research on geoengineering has been already carried out \cite{vaughan2011review,macmartin2018solar,shepherd2012geoengineering}: devising an optimal strategy to achieve specific climate targets using closed-loop control via solar radiation modification (SRM) for attaining multiple climate targets \cite{kravitz2016geoengineering,macmartin2013management,macmartin2014dynamics}, and using linear response theory to estimate the optimal spatial pattern of SRM to offset the climatic effects of greenhouse gas warming \cite{bodai2020can}, for example. Some studies have also focused on the potential ramifications of geoengineering strategies such as inadvertent regional anomalies in temperature and precipitation \cite{schneider2008geoengineering,irvine2016overview,lawrence2018evaluating,kravitz2011geoengineering} and socio-political fallout \cite{abatayo2020solar, moore2021targeted}.

 However, to the best of our knowledge, there are currently no studies that address the question of controlling climate in the vicinity of a tipping point. Past research on climate tipping points has mainly focused on using early warning signals to anticipate tipping points \cite{lenton2011early, scheffer2012anticipating, lenton2015detecting,bury2021deep} and estimating time of approach \cite{ditlevsen2023warning, hankel2023approach}. Furthermore, the existing studies on the design of climate control are based on achieving only a small number of global-scale climate targets through feedback control with limited degrees of freedom \cite{kravitz2016geoengineering,kravitz2017first} or through spatially-varying passive control using linear perturbation theory \cite{bodai2020can}. 

 The present study aims to bridge these gaps by estimating and analyzing a spatiotemporal, closed-loop, \add{albedo-based} control strategy that is optimal with respect to a prescribed cost function, for stabilizing or reversing sea-ice loss in the vicinity of a tipping point. To this end, we use a simple energy balance model (EBM) to represent climate dynamics.  This EBM has a tipping element that arises from the strong nonlinearity associated with the increased reflectivity of sea-ice or ice-albedo \cite{held1974simple, north1981energy}. The simplicity of the EBM allows us to employ an analytical approach by framing the question of control as a PDE-constrained optimization problem \cite{hinze2008optimization}. 
 
 We succeed in devising an optimal control scheme to stabilize sea-ice while minimizing the cost of applying this scheme, by employing variational methods \cite{strang1986introduction}. Essential features of the tipping point, such as the critical threshold and the strength of hysteresis, can be quantified and tuned by varying the parameters of the EBM. This allows us to study the dependence of control measures on tipping point attributes. We find that the EBM can be controlled around the sea-ice tipping point and that the control is most cost-effective when applied near the ice-edge. \add{We show that the control intervention required to reverse sea-ice loss nearly doubles once the system starts undergoing a rapid transition to the ice-free state or `tipping'. We also explore the possible delay in the onset of tipping or an overshoot past the tipping threshold \cite{ritchie2021overshooting, bathiany2018abrupt} due to system inertia and how this affects the scaling of control costs with the time of deployment. An increase in the system inertia allows for a larger time window for intervention during which the control cost scales linearly with the intervention time; past this limit, there is a steep rise in the costs. And our studies show that the larger this leeway, the larger the cost of reversing sea-ice loss once the window is crossed.}
 Furthermore, the spatiotemporal maximum of control required in the post-tipping scenario is an order of magnitude higher than that of the pre-tipping case. These results have important implications for intervention strategies and policy design related to the prevention of catastrophic climate change.

\section{EBM: Bifurcations and Tipping Points}

The EBM was developed in the classic papers of Budyko and Sellers \cite{budyko1969effect, sellers1969global}. Here, we use a standard one-dimensional diffusive EBM \cite{north1981energy}; it is derived by considering the heat balance on an infinitesimal strip centered at latitude $\theta$. In our EBM, the zonally-averaged, annual mean surface temperature $T$ evolves as
\begin{align}
    C \frac{\partial T}{\partial t} =& \;(1-\alpha(T))Q_{SW}(x) - (aT + b_0) + F(t) + D\frac{\partial}{\partial x} \left( (1-x^2) \frac{\partial T}{\partial x} \right) 
\end{align}
where $x=\sin{\theta}$ is the latitudinal coordinate, and subject to boundary conditions imposing no heat flux at the pole and equator. For simplicity, the system is assumed to be symmetric about the equator and therefore the spatial domain considered is $ 0^\circ \le \theta \le 90^\circ$. The ocean mixed layer heat capacity is given by $C$ and is assumed to be a constant. The latitude-dependent incoming annual mean shortwave (SW) radiation is $Q_{SW}(x)$, the outgoing longwave radiation (OLR) term is $aT + b_0$ \cite{north1975analytical} and the external forcing applied to the EBM to mimic the effect of increasing/decreasing GHG concentration is given by $F(t)$. The ice-albedo $\alpha(T)$ is temperature dependent and is assumed to be a smoothed step function with $\alpha = 0.32$ for $T \gg -10^\circ$ C and $\alpha = 0.62$ for $T \ll -10^\circ$ C; these asymptotic albedo values are based on the Budyko formulation in \cite{budyko1969effect}. A more detailed discussion of the EBM used in this study can be found in appendix \ref{sec:A}, and the EBM parameter values are given in Table \ref{tab:EBM_params}.

The sharp transition in ice-albedo across the ice edge, taken to be at $T = -10^\circ$ C, results in a strong positive nonlinear feedback with important consequences for the dynamics of this model. The EBM with albedo feedback is known to exhibit multistability \cite{north1990multiple}, which is defined as the existence of multiple possible time-invariant states for a given set of parameters. \add{Continuous changes in system parameters produce qualitative changes in the stability characteristics of the EBM.  These changes are known as bifurcations, and the existence of these bifurcations in the EBM is what gives rise to multistability (for certain values of those system parameters)} More specifically, all the bifurcations are of saddle-node type, characterized by the appearance or disappearance of a pair of steady-state solutions \cite{ghil2012topics, wigginsintroduction} as a system parameter is varied. Saddle-node bifurcations cause catastrophic shifts and are associated with hysteresis behavior; many other tipping points of the Earth system such as AMOC and West Antarctic Ice Sheet are also \add{hypothesized to be} associated with saddle-node bifurcations \cite{ditlevsen2023warning, rosier2020tipping}. 
 
In the EBM, \add{the system parameter creating these bifurcations is $F\left(t\right)$ (i.e., the level of GHG forcing).  T}he runaway feedback from the strongly nonlinear albedo during a global cooling event can lead to a sudden transition to the so-called `snowball' Earth \cite{pierrehumbert2011climate} once the ice cover crosses a critical latitudinal threshold. This tipping point is commonly referred to as a \emph{large icecap} instability. 

A second tipping point is a \emph{small icecap} instability (SICI) \cite{north1984small,north1990multiple} which leads to irreversible loss of icecaps smaller than a characteristic length scale associated with diffusive energy transport. The bifurcation curve \add{corresponding to the SICI} (dotted line) in Fig \ref{fig:bt_response}\add{b} represents the steady solutions to the EBM for different values of the control parameter $F$ (refer to appendix \ref{sec:C} for computational details). Three different branches of steady solutions can be noted: the top branch, with the ice-line at the pole, corresponds to the ice-free stable solutions; the bottom branch consists of stable solutions with a finite icecap; and the middle branch represents the intermediate unstable solutions. The yellow diamonds represent the two saddle-node bifurcation points associated with SICI. The right bifurcation point represents the limit past which the EBM has no finite icecap steady solutions. Similarly, the left bifurcation point represents the end of the ice-free branch. The small icecap instability has been a subject of intense scrutiny over the past decades owing to its implications for a rapid, irreversible transition to a perennially ice-free Arctic under greenhouse warming \cite{eisenman2009nonlinear, hezel2014modeled}. 

\subsection{Response to greenhouse warming}
\begin{figure}[t!]
\centering
\includegraphics[width=0.7\linewidth]{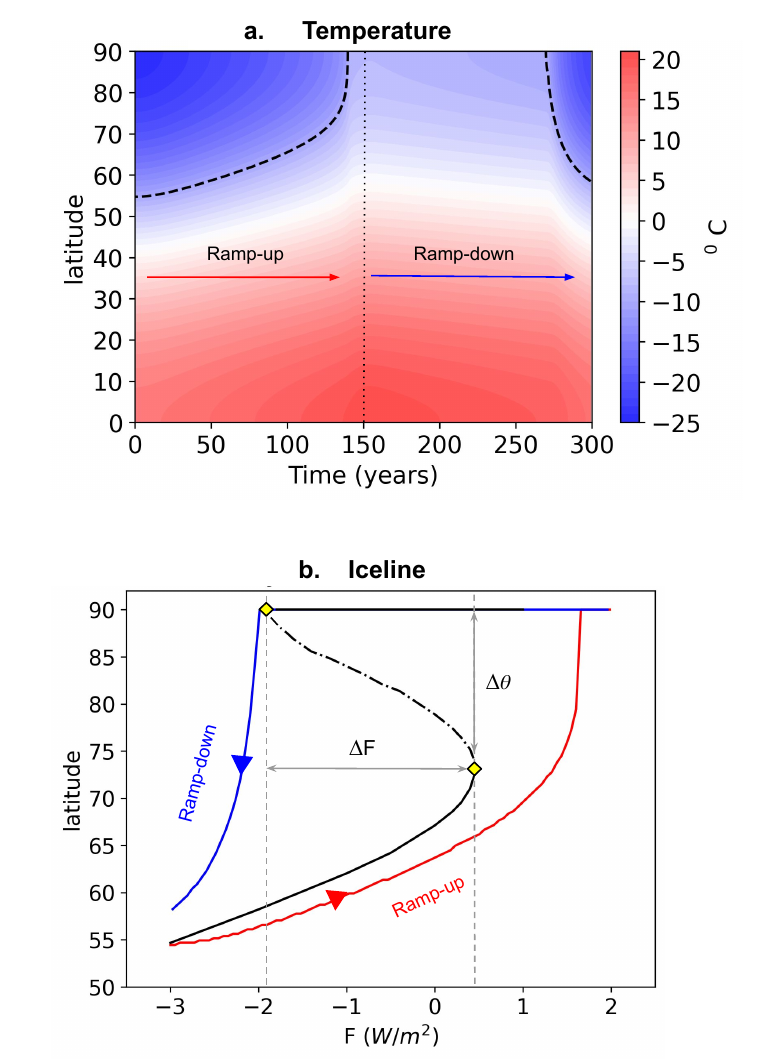}
\caption{\add{a.} Evolution of the temperature profile and ice-line (black dashed line) in response to a spatially uniform, linear ramp-up and ramp-down forcing applied over a period of 300 years. \add{b.} The solid line is the hysteresis curve for the ice-line under the applied forcing. The dotted line represents the bifurcation diagram for SICI and the yellow diamonds are the two saddle-node bifurcation points.}
\label{fig:bt_response}
\end{figure}
 In order to understand the response of the EBM to greenhouse warming in the vicinity of the SICI, we subject it to a spatially uniform, linear ramp-up in the GHG forcing parameter $F$ such that the \add{increase in} forcing is 5 W$/$m$^2$ at the end of 150 years \add{($\approx$ 0.033 W/m$^2$/yr)} and is then brought back linearly to \add{$F = -3$ W/m$^2$} over the next 150 years. The starting state is the steady solution corresponding to $F = -3$ W/m$^2$.

In the presence of the forcing defined above, the evolution of the temperature profile and ice-line, indicated by the black dashed line, is shown in Fig.\ref{fig:bt_response}\add{a}. The irreversibility associated with SICI manifests as asymmetry in the responses of mean temperature and ice-line during the ramp-up and ramp-down phase. Initially, the icecap recedes linearly; however, around year 140, there is an almost abrupt disappearance of sea-ice cover, which does not recover until year 270. 

The hysteresis curve (solid line) corresponds to the time-varying solutions in response to the applied forcing and is shown in Fig. 
\ref{fig:bt_response}\add{b} along with the bifurcation diagram for this model. It does not follow the bifurcation curve (dotted line) exactly because the rate at which the applied forcing varies is too rapid for the system to adjust to steady state. Consequently, the bifurcation diagram only provides an estimate of when the system is likely to tip. \add{The tipping time in real-world systems is determined by the interplay of the short-term variability in the climate system (modeled as noise to the long-term variability of climate) and system response time. As the system approaches the tipping threshold, the restorative forces that damp perturbations weaken (critical slowdown \cite{ditlevsen2010tipping,lenton2011early}), and the susceptibility to noise increases - a strong perturbation may therefore cause the system to tip well before the forcing threshold \cite{ashwin2012tipping}. On the other hand, if the external noise levels are low, the tipping time will be determined by the system response time. Owing to the system inertia, the tipping threshold may be exceeded temporarily without triggering rapid transition, referred to as `overshooting' \cite{bathiany2018abrupt,ritchie2019inverse, ritchie2021overshooting}.}

\add{In Fig. \ref{fig:bt_response}b}, we define the hysteresis width $\Delta F$ as the horizontal distance between the bifurcation points (yellow diamonds) and can be used \add{as} a measure of the degree of irreversibility and instability \cite{wagner2015climate}. The tipping size $\Delta \theta$ is defined to be the vertical distance between the bifurcation points and can be interpreted as the critical size for receding icecaps, past which the system transitions rapidly to an ice-free state in a warming climate.  During the ramp-up (red line), the system initially remains close to the stable finite-ice branch but as it approaches the tipping point, there is accelerated loss of ice and the system quickly transitions to the ice-free branch, i.e., ice-line retreats to the pole. During much of the ramp-down stage (blue line), the system remains on the ice-free branch and it is only once it crosses the second saddle-node bifurcation that the system recovers sea-ice. The hysteretic behavior implies that once the tipping point is crossed, modest efforts to reduce forcing alone will not suffice to reverse sea-ice loss. 

\section{Materials and Methods}

An effective control strategy in this context is one that can arrest or even reverse the crossing of the small icecap tipping threshold. Such a control strategy must be amenable to the strong nonlinearity near the critical threshold and adaptable to the different regimes of the system resulting from multistability.  

\add{The control mechanism chosen for our study is an artificially introduced planetary albedo-based control $u(x,t)$, that can vary based on latitude. Note that estimates of $u(x,t)$ can be used to design any general solar radiation management methods that locally modify the planetary albedo such as stratospheric sulfate aerosol injection \cite{lee2023high}, marine cloud brightening \cite{latham2014marine} and surface albedo modification \cite{field2018increasing}.} Once we solve for the optimal albedo control, the optimal external forcing required for a general control strategy can be obtained as $uQ_{SW}$. The controlled EBM is then given by,
  \begin{align}
    C\frac{\partial T}{\partial t} &= (1-\alpha(T) - u(x,t))Q_{SW}(x) - (aT+ b_0)  + F(t) + D\frac{\partial}{\partial x} \left( (1-x^2) \frac{\partial T}{\partial x} \right).
    \label{eqn:EBM_ctrl} 
\end{align}  

Note that there exists a control $u_e(x,t)$ that exactly cancels any forcing $F(x,t)$ at all times $t$:
\begin{align*}
   u_e(x,t)= \frac{F(x,t)}{Q_{SW}(x)}.
\end{align*}

 For a spatially uniform forcing, the requisite exact control $u_e$ increases monotonically from the equator to the pole. Solar insolation drops as we move away from the equator, so at higher latitudes, a larger fraction of the incident radiation needs to be reflected to offset the same forcing (see SI for details).

 Exact control, being passive, does not alter the stability characteristics of the EBM and therefore is ineffective once the \add{sea-ice extent drops below the critical value (tipping size)}. Additionally, exact control does not take into account the cost associated with the deployment of control, so it is not cost-efficient.

\subsection{Optimal control} 
We first formulate a cost function that accounts for the deviation from some target temperature profile $T_f(x)$ and the scale of intervention $u(x,t)$ at every time $t$ and spatial location $x$. Here, the cost function $\mathcal{C}$ is chosen as
\begin{align}
    \mathcal{C}(T(x,t),u(x,t)) =  \left( w_u \frac{u^2}{\bar{u}^2} + w_T \frac{(T - T_f)^2}{\bar{T}^2} \right) 
    \label{eqn:cost_functional}
    \end{align}
where $T_f(x)$ is the desirable state to be attained and the deviation from the target profile and the control are suitably normalized by $\bar{T}$ and $\bar{u}$, respectively (see appendix \ref{sec:B}). The relative prioritization of control cost and temperature deviation, which will depend on the climate policy goals, will be reflected in the relative values of the weights $w_u$ and $w_T$. The objective can then be written as

\begin{align}
\min_{u(x,t)} \mathcal{L}[u, \; &T] = \min_{u(x,t)}  \int_0^{t_f} \int_0^1 dx \; dt \; \mathcal{C}(T(x,t),u(x,t)) 
\label{eqn:min}
    \end{align}
where $\mathcal{L}[u, T]$ is the total cost functional, and $t_f$ is the time period over which the control is to be applied. We also need a constraint on $u$ to ensure that the net albedo remains less than one for all $x$ and $t$ -- i.e.,
$$0\le u + \alpha \le 1.$$
This constraint is imposed implicitly by restricting the weight ratio $w_T/w_u$. As the relative penalty on the control decreases, the fraction of the cost associated with control in equation (\ref{eqn:cost_functional}), increases. Therefore, by keeping $w_T/w_u$ under some upper bound $\lambda_{max}$, the constraint can be satisfied. This upper bound depends only on the EBM parameters and can be determined numerically.
 
The principles of variational calculus can be employed to derive the Euler-Lagrange equation  \cite{strang1986introduction} corresponding to the minimization problem in equation (\ref{eqn:min}), and is given by
\begin{align}
    \frac{u}{\bar{u}^2} + \frac{w_T}{w_u}\left( \frac{T - T_f}{\bar{T}^2 } \right) T_u = 0.
    \label{eqn:ebm_opt_1}
\end{align}
Therefore, at every time $t$, the optimal control $u$ can be computed using the instantaneous $T$ \add{from (\ref{eqn:EBM_ctrl})} and the local sensitivity to control, $T_u := \frac{\partial T}{\partial u}$.  Consequently, to compute $u(x,t)$, $T_u(x,t)$ must be known and this can be done by deriving an additional equation for the time evolution of $T_u$ from the EBM. \add{This system of PDEs for the controlled EBM is then solved by integrating forward in time.} A detailed derivation of the optimal control can be found in appendix \ref{sec:B} and \add{the computational methods are described in appendix \ref{sec:C}}.

The weight ratio $\frac{w_T}{w_u}$ is a free tuning parameter to balance the cost of control and the rate of approach to the target temperature profile. This will be explored in detail in a subsequent section.

\section{Results}

\begin{figure*}[t!]
\centering
\includegraphics[width=\linewidth]{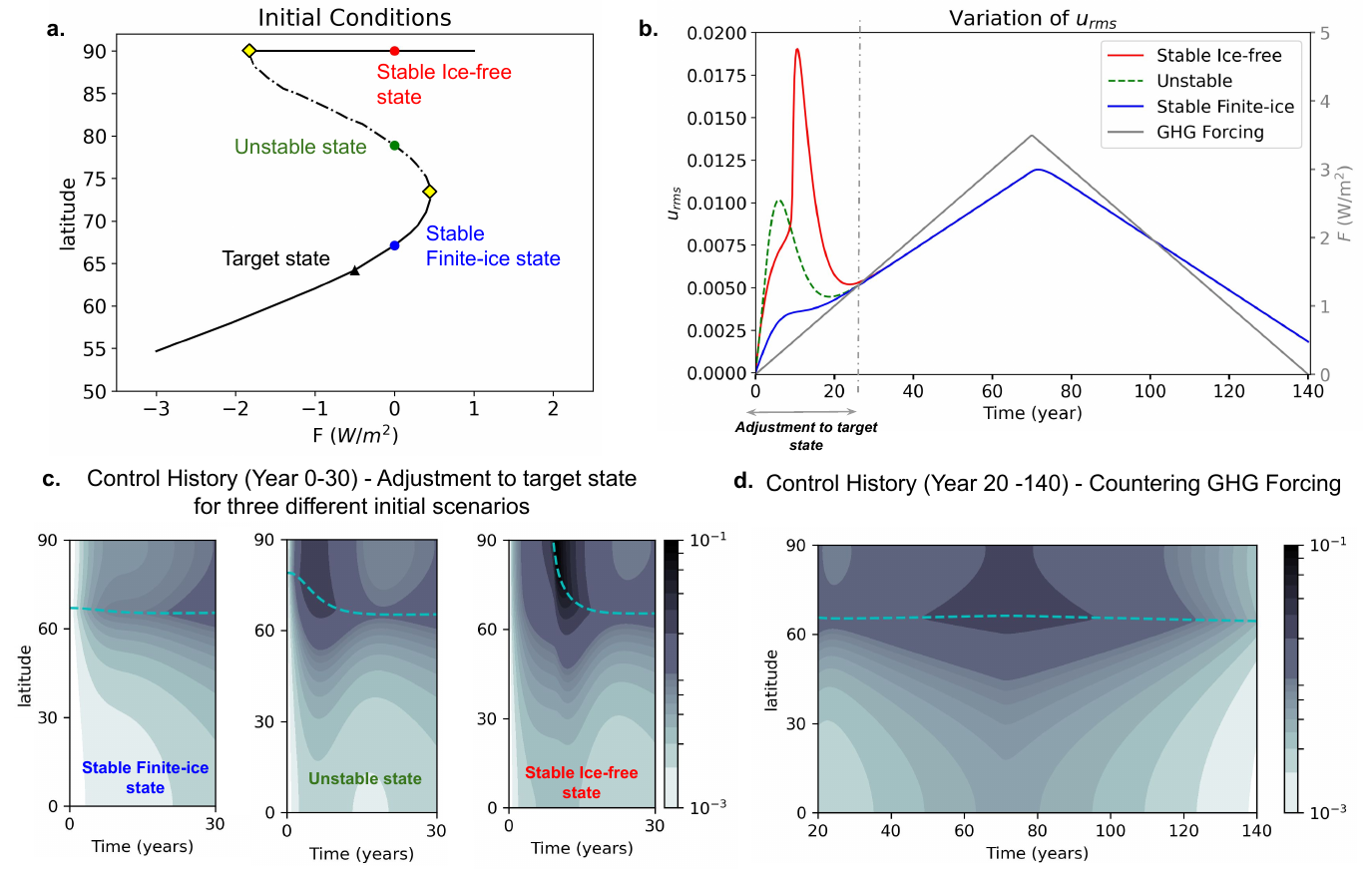}
\caption{\add{Case 1: }Three scenarios for optimal control. Panel a: Bifurcation diagram indicating the location of the three initial states and the common target state. Panel b: Variation in optimal control costs ($u_{rms}$) with time for the three scenarios employing the same control strategy with $\frac{w_T}{w_u} = 0.1$. Also shown is the applied GHG forcing (grey curve). Panel c: Contour plots of optimal control $u$ for the initial adjustment to the target state. Panel d: Contour plot of control $u$ after year 20.}
\label{fig:start}
\end{figure*}

 \add{In what follows, we consider scenarios relevant to two plausible tipping pathways - one in which the system tips prematurely and the other in which tipping is delayed due to system inertia or overshooting.}
\add{\subsection{Case 1}
 The effects of internal variability and other sources of noise on the control are neglected here for simplicity. However, while examining the possible tipping scenarios, one must take into account the role of external noise in determining the tipping time - a system that is `close' to its bifurcation point may tip prematurely if the noise intensity is strong enough to cross the potential barrier height associated with the attractor basin for the steady solution \cite{ditlevsen2010tipping, ritchie2019inverse}, i.e., the system may transition to the ice-free regime before the GHG forcing exceeds the tipping threshold due to the destabilizing effects of noise.}

\add{In order to probe this tipping pathway,} three different starting scenarios are considered for optimal control. Fig. \ref{fig:start}a shows the starting states on the three branches of the bifurcation diagram associated with SICI, all of which correspond to $F=0$ W/m$^2$. Two of these states are stable (ice-free and finite-ice)\add{; the ice-free state represents a plausible final state following a perturbation-induced transition}. \add{We also choose a point on the unstable equilibrium line as it provides a convenient reference point.  It could, for example, be a point on a dynamic non-equilibrium trajectory that undergoes a premature transition.} The unstable starting state allows us to investigate the efficacy of the optimal control in stabilizing and reverting the system to a stable finite-icecap state -- even in the presence of GHG forcing. The chosen target state (filled-in triangle) corresponds to $F = -0.5$ W/m$^2$ on the finite-icecap solution branch. The EBM is subject to a spatially uniform linear ramp-up and ramp-down forcing (up to $3.5$ W/m$^2$ and back to zero) over a period of 140 years, mimicking the ramp-up of CO$_2$ forcing at a rate of 1$\%$/year followed by a ramp-down at the same rate to the original CO$_2$ forcing level (Fig. \ref{fig:start}b) \add{that represents the gradual lowering of CO$_2$ to pre-industrial levels through the lowering of emissions, carbon capture, etc \cite{cao2023simulated}. The model itself does not require us to specify the mechanism by which the forcing level changes, so it is sufficient for our purposes here that there exist plausible examples of such mechanisms.} The weight ratio $w_T/w_u$ is set to be $0.1$ for which the applied albedo control is effective and remains physical.

For each of the three scenarios, the root-mean-square (RMS) of control at time $t$, a measure of the cost of that control (see equation (\ref{eqn:cost_functional})),
\begin{equation}
   {u}_{rms} \left(t\right) = \sqrt{\int_0^1 dx \; {u^2 \left(x,t\right)}}, \nonumber
\end{equation}
is compared in Fig. \ref{fig:start}b. 

The response of the EBM with control can be roughly divided into two segments. The first stage ($t \le $ 35 years) is dominated by the initial transition via control to the target state; the differences in the requisite control amongst the three starting conditions are manifested mainly during this time interval. The time scale for this adjustment is set by the ocean heat capacity $C$ and the weight ratio $w_T/w_u$ which constrains the growth of control. Reducing $w_T/w_u$ lowers the magnitude of control $u$ during the adjustment to the target state. 

Significant differences can be noted in the cost associated with the initial adjustment for the three scenarios. The area under the RMS control curve in Fig. \ref{fig:start}b, during the adjustment phase for the unstable starting state is approximately twice that for the finite-ice starting case and the corresponding control cost, $u_{cost}$ for the first 20 years is nearly four times that of the unstable case.
From Fig. \ref{fig:start}c, it is clear that most of this increased cost can be attributed to the transition from the unstable branch to the stable finite-ice branch. For the initial state in the ice-free regime, the maximum of RMS control for the adjustment is about four times that of the control required for the finite-ice initial state. However, $u_{rms}$ does not tell us the whole story: the maximum local albedo control required for the stable finite-ice adjustment is 0.01 while those for the unstable and stable ice-free states are approximately 0.03 and 0.1, respectively, as seen in the three sub-panels in Fig. \ref{fig:start}c. Therefore, the restoration of sea-ice starting from the ice-free branch requires a large external albedo imposed over a sizeable latitudinal zone and this will likely have significant ramifications for the global climate. Another characteristic of note is that for the case starting from a stable ice-free state, there is a sharp increase in applied control once the ice reappears at around year 10 and lasts until the target ice-line is achieved. \add{The optimal control given by (\ref{eqn:ebm_opt_1}) depends on both the deviation from the ideal temperature profile and the sensitivity of temperature to control $T_u$. The reappearance of ice bolsters the sensitivity to control $T_u$ since the local positive ice-albedo feedback now works in synergy with the external albedo $u$ to reduce $\Delta T_f$ and results in the sharp increase of $u$ near the ice-line to take advantage of the increased sensitivity there.} Note that the time scale associated with ice reappearance is set by the choice of weight ratio $w_T/w_u$, which regulates the rate at which the applied control increases, and the heat capacity $C$, which modulates the temperature response rate.

The second stage corresponds to the stabilized stage when the optimal control is identical for all three scenarios and only works to counteract the ramp-up and ramp-down GHG forcing to maintain the target state. A key feature of the optimal control in both stages is the spatial localization around the ice-line: the largest values of $u$ are along the ice-line. This is in contrast to the exact control, which requires maximum albedo change at the poles. This observation suggests that spatially localized control interventions may be adequate for reversing the rapid loss of sea-ice.

\add{\subsection{Case 2}}

\begin{figure*}[t]
\centering
\includegraphics[width=\linewidth]{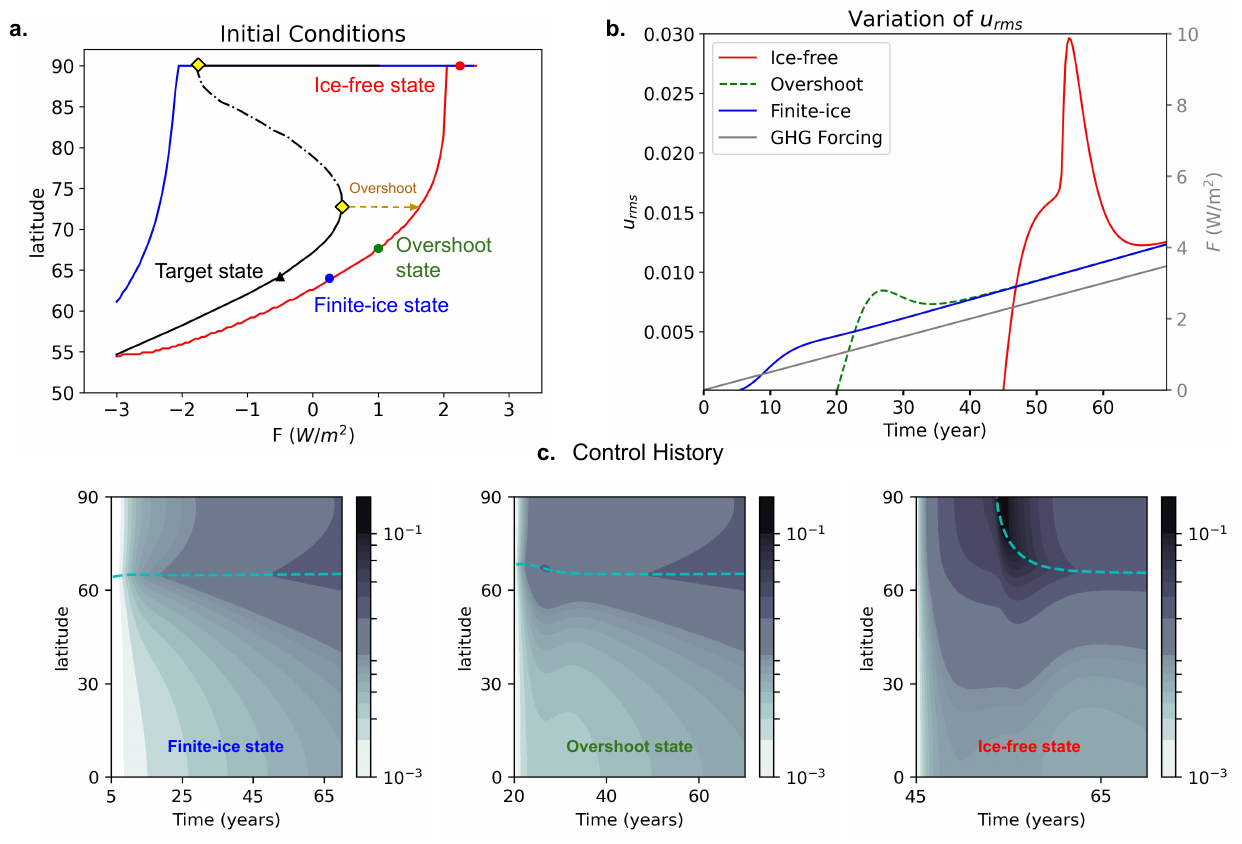}
\caption{\add{Case 2: Three scenarios for optimal control. Panel a: Bifurcation diagram and hysteresis curve indicating the location of the three initial states and the common target state. Panel b: Variation in optimal control ($u_{rms}$) with time for the three scenarios employing the same control strategy with $\frac{w_T}{w_u} = 0.1$, but with different times of intervention. Also shown is the applied GHG forcing (grey curve). Panel c: Contour plots of optimal control $u$.}}
\label{fig:start2}
\end{figure*}

\add{In this section, we explore intervention scenarios that account for the possible overshoot of the tipping threshold due to system inertia, which is regulated by the heat capacity $C$ in the EBM. To understand the dependence of the control requirement on the system response time, three out-of-equilibrium initial conditions are considered on a time-varying trajectory of the EBM. The system is allowed to evolve from the steady state at $F=-3$ W/m$^2$ subject to ramp-up forcing (0.05 W/m$^2$/yr) and three different intervention times are chosen as shown in Fig. \ref{fig:start2}a. The first corresponding to $F=0.25$ W/m$^2$ that is just before the tipping threshold ($F = 0.45$ W/m$^2$). The second initial condition is chosen at $F = 1$ W/m$^2$, which is in the overshoot phase.  The final one is chosen at $F=2.25$ W/m$^2$, by which time the transition to the ice-free regime is complete. The target state is chosen to be the same as in case 1.}

\add{The variation in $u_{rms}$ for the three intervention times are shown in Fig \ref{fig:start2}b. Similar to case 1, the major differences between the three intervention times occur during the adjustment to the target state. All three states evolve identically after the target state is attained. Therefore, the appropriate measure to compare the cases is the additional control requirement associated with the delayed intervention ($\int \Delta u_{rms} dt$) and is defined here as the area enclosed by the closed loop consisting of the $u_{rms}$ curve for delayed intervention (green or red) and the pre-tipping (blue) control curve. As the time of intervention is delayed, this additional control requirement increases. In comparison with the overshoot case, $\int \Delta u_{rms} dt$ for the ice-free starting state is significantly larger.}

\begin{figure}[b]
\centering
\includegraphics[width=0.9\linewidth]{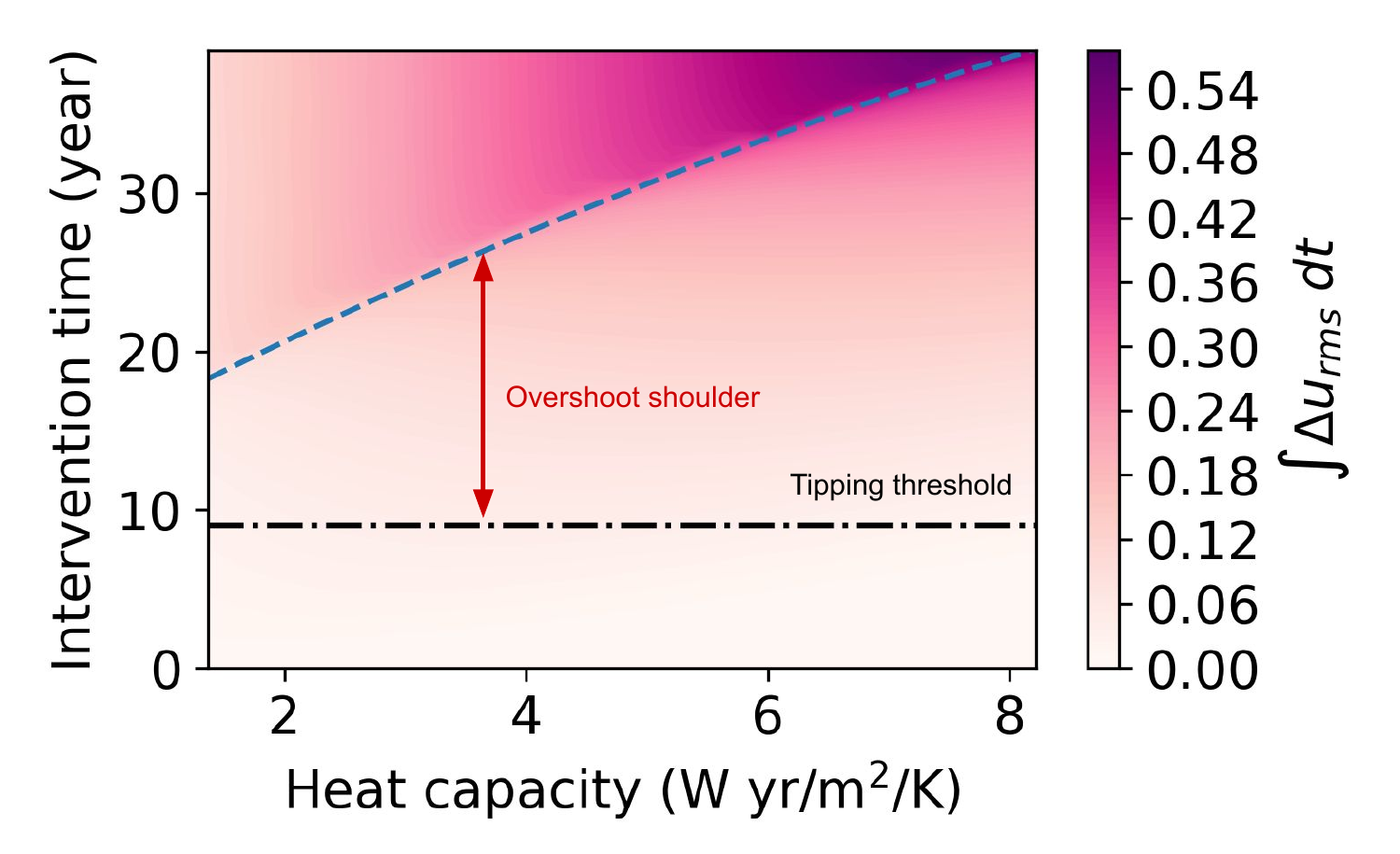}
\caption{\add{Contour plot showing the variation in control requirement as a function of heat capacity $C$ and the time of intervention for case 2. The dashed line represents the time of intervention at which there is a steep increase in the requisite control which can be approximated by a quadratic fit given by $0.05t_o^2 + 1.3t_o \approx 10C$; the dash-dotted line represents the time at which the GHG forcing crosses the tipping threshold.}}
\label{fig:u-C-ti}
\end{figure}

\add{To further explore the role of the overshoot in determining the control requirement, the above analysis is repeated for a range of intervention times, $0 \le t_i \le 40$ (in years), and for a range of heat capacities, $1.4 \le C \le 8.2$ (in W yr/m$^2$/K). The additional control requirement ($\int \Delta u_{rms} dt$) as a function of heat capacity and intervention time is shown in Fig \ref{fig:u-C-ti}. The time ($t= 9$ years) at which the ramp-up forcing reaches the tipping threshold is indicated by a dash-dotted line. For a given heat capacity, as the intervention time exceeds a certain limit, an almost step change in the control requirement can be noted. The dashed line in Fig \ref{fig:u-C-ti} represents a simple quadratic fit for this limiting intervention time past which there is a steep rise in the control requirement. The extra time afforded by the system response time scale or the `overshoot shoulder' ($t_o$) is defined as the distance between the two lines. As the system response becomes sluggish, there is a larger leeway in the intervention time, but this is accompanied by a steeper change in the control requirement past the overshoot shoulder. Therefore, even though increased system inertia delays tipping, that same system inertia significantly increases the control costs once the system is past the overshoot shoulder; the cost increase, in and of itself, is not surprising, but the suddenness of that increase beyond a certain time point is unexpected.} 

\subsection{Tuning weight ratio for optimal control}

\begin{figure}[h]
\centering
\includegraphics[width=0.7\linewidth]{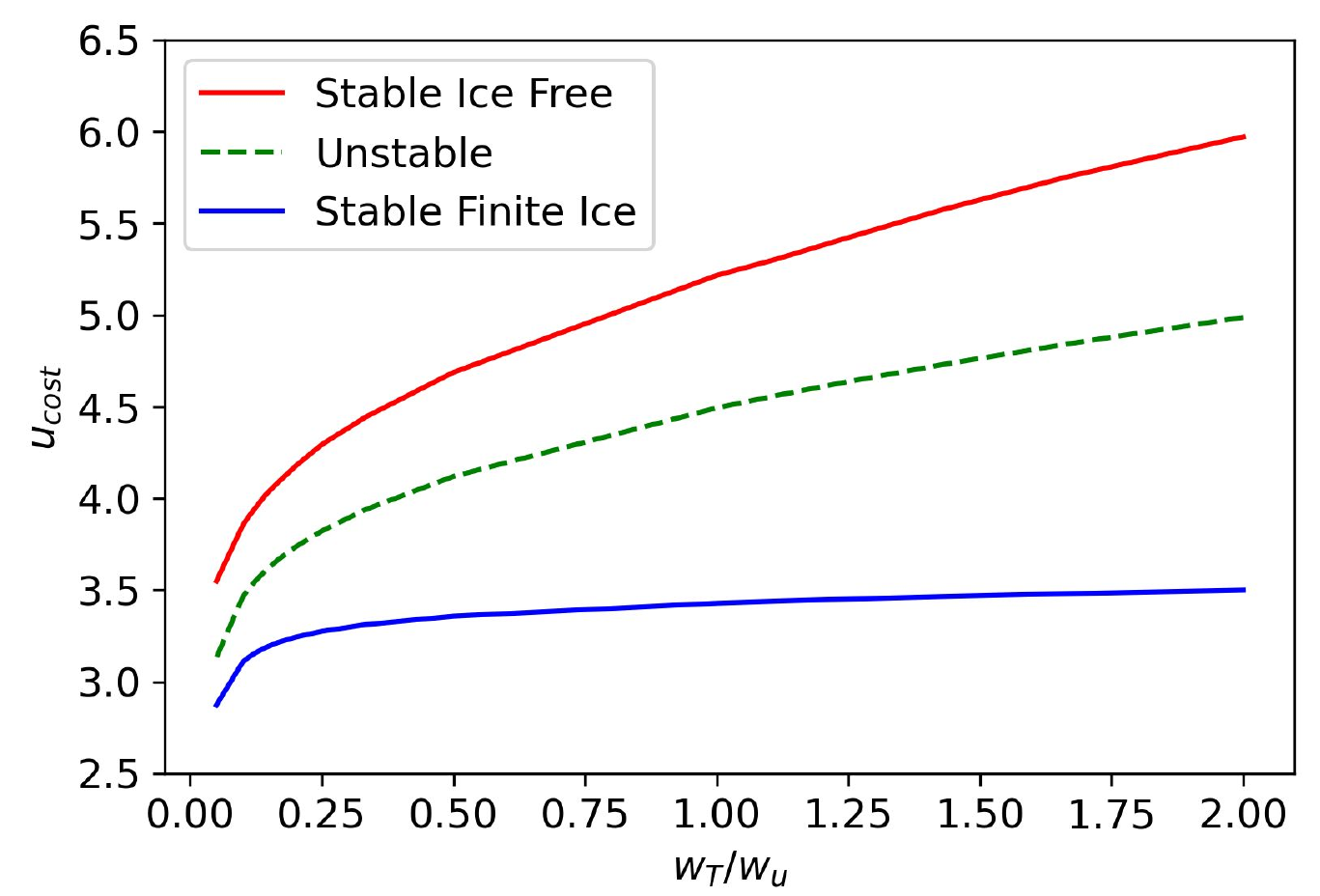}
\caption{Variation in cost of control as a function of weight ratio}
\label{fig:weight_ratio}
\end{figure}

In this formulation, and given the parameter values in Table \ref{tab:EBM_params}, the weight ratio ${w_T}/{w_u}$ has a bounded operational range. For values of ${w_T}/{w_u}$ less than $\lambda_{min}\approx 0.04$, the control, $u$ does not scale up fast enough to effectively control the climate. For values of $w_T/w_u$ greater than $\lambda_{max} \approx 6.7$, the control $u$ can produce a net albedo greater than one (which is unphysical).

For the three initial conditions described above, the cost of control,
\begin{equation}
   u_{cost} = \int_0^T \int_0^1 dt \; dx \; {u^2} \nonumber 
\end{equation}
varies with weight ratio within the operational range, and is shown in Fig. \ref{fig:weight_ratio}. From Fig. \ref{fig:weight_ratio}, it is clear the control costs for all three states increase when the deviation from the target temperature profile is given more weight than the applied control, as one would expect. The differences in costs between the three initial states also increase with larger $w_T/w_u$ values -- especially near the upper limit. Also noteworthy is the almost constant cost for the pre-tipping scenario (blue curve). The weak sensitivity of the total control cost for the case of the pre-tipping control implies that one may prioritize, and thus hasten, the attainment of the climate target without significantly increasing the cost of control. The results in Fig. \ref{fig:start} compare the control costs for the three scenarios corresponding to $w_T/w_u = 0.1$, which is close to the lower limit of the operational range (which is approximately 0.04, as mentioned previously) and therefore provides an approximate lower bound for the control required around the SICI.

\subsection{Tipping point characteristics}

 Hysteresis width $\Delta F$ is defined as the width of the hysteresis loop and has units of $W/m^2$ as shown in Fig. \ref{fig:bt_response}. Hysteresis width can be used to quantify the irreversibility associated with a tipping point and may be used to characterize the behavior of a tipping point. The hysteresis width is mainly determined by the nonlinearity in the co-albedo and the diffusivity parameter $D$ in the EBM.  The latter can be tuned to vary the hysteresis width associated with SICI \cite{wagner2015climate}.

\begin{figure}[h]
\centering
\includegraphics[width=0.8\linewidth]{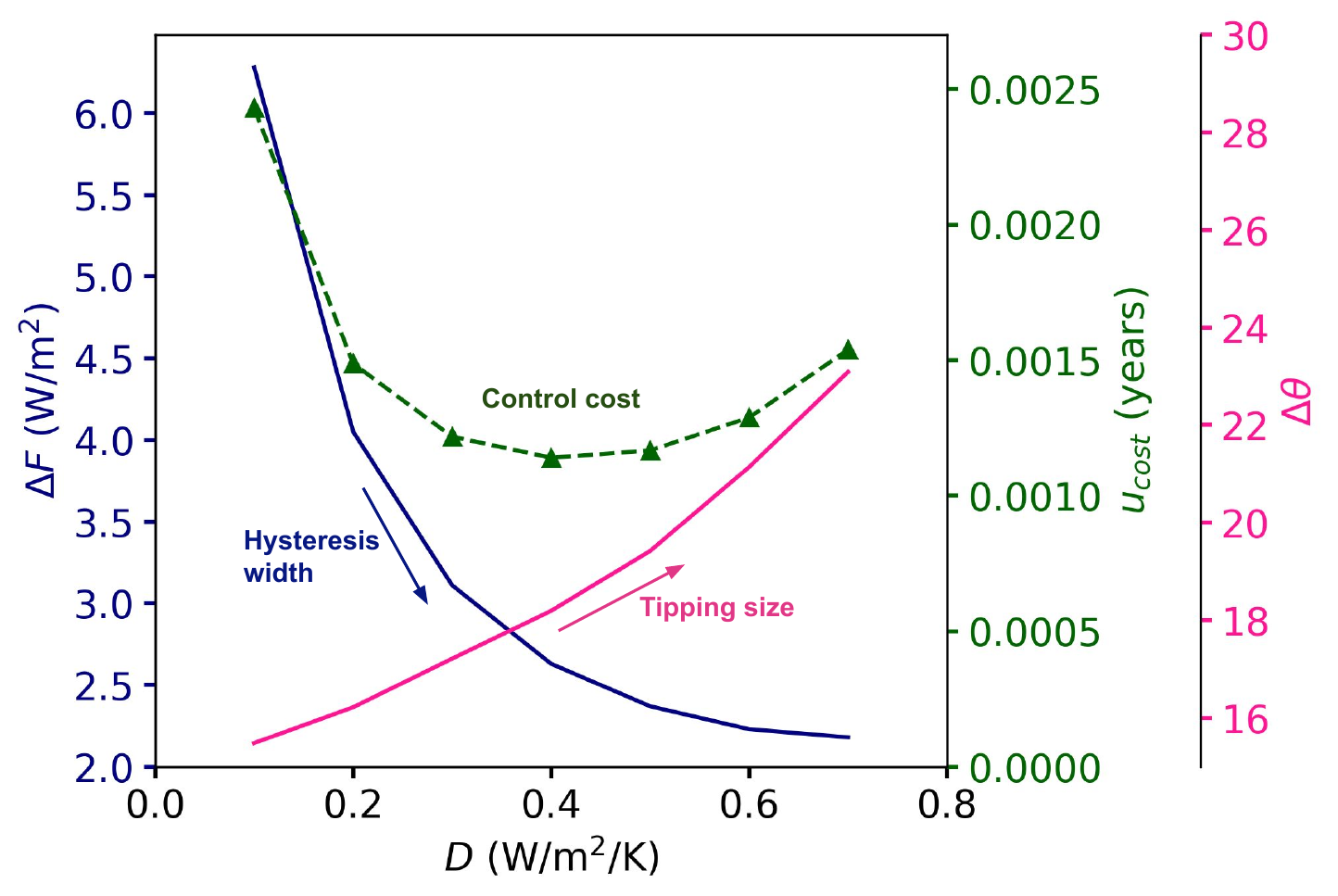}
\caption{Variation in hysteresis width ($\Delta F$), tipping size ($\Delta \theta$) and cost of control ($u_{cost}$) with diffusion coefficient ($D$)}
\label{fig:u_D_dF}
\end{figure}

To understand how control costs scale with hysteresis width, we first compute the control cost of restoring the finite-ice stable state from the ice-free initial state at a distance of $0.1 \Delta F$ from the SICI tipping point (similar to the red and blue dots in Fig. \ref{fig:start}a, which are at a distance of approximately $0.2 \Delta F$) and then compare control costs as $D$ and the associated $\Delta F$ systematically vary. \add{This study is carried out at constant GHG forcing in order to isolate and analyze the cost of adjustment from the ice-free regime to the finite-ice regime.}  Intriguingly, the control costs (with $w_T/w_u = 0.5$) do not scale monotonically with hysteresis width (Fig. \ref{fig:u_D_dF}). Rather, the costs initially decrease as the hysteresis width decreases with increasing $D$. Then, upon reaching a minimum at around $D=0.4$, the costs start to increase. This behavior can be understood once the tipping size $\Delta \theta$ (see Fig. \ref{fig:bt_response}) associated with the SICI---the distance between the latitude of the tipping point and the pole---is taken into account. The tipping size provides a measure of how much ice growth is required before the system will self-adjust to the stable finite-ice branch. The tipping size is known to \add{increase} in an EBM with increasing diffusivity (see \cite{north1990multiple}) and this is responsible for the increase in costs at higher diffusivities.  Significant uncertainties exist in our estimates of critical thresholds and irreversibility for tipping elements \cite{armstrong2022exceeding, wang2023mechanisms}, so it is imperative to consider the effect of these attributes when evaluating system control characteristics and reversibility around a tipping point. \add{We also carried out an analogous study that considers the scenario in Case 2 subject to a linear ramp-up in GHG forcing and these qualitatively similar results are shown in the SI. Additionally, the control sensitivity to the nonlinearity in the co-albedo is also explored in the SI.}

\section{Discussion}


We designed and analyzed an optimal control strategy in the vicinity of the small icecap tipping point in an idealized climate model. This is a first attempt at understanding the first-order features of optimal control around climate tipping points. A key insight from this analysis is the \add{quantitative} confirmation that preventative measures for averting tipping are much less intrusive and less costly than corrective post-tipping interventions to restore the climate state. \add{Our study shows that overshoot due to system inertia may afford a finite intervention window past the tipping threshold during which the control costs only scale linearly with the delay in intervention. And as the system response time increases, the length of this intervention window increases. Crucially, our results reveal that this is no free lunch -- this increased leeway is accompanied by a considerable increase in the intervention costs once the overshoot window is crossed.} Furthermore, the relatively weak dependence of the total control cost on the weight ratio $w_T/w_u$ (Fig. \ref{fig:weight_ratio}) implies that for early intervention before tipping, it is possible to effect a faster reversal of sea-ice loss and thus a closer adherence to the climate target without significant increase in control costs.
  
 For the sea-ice tipping, the spatial distribution of the optimal control suggests the localized geoengineering measures, in this case, restricted to the neighborhood of the ice-line, are most impactful in achieving requisite control of the tipping element. We also showed that characteristics of the sea-ice tipping element such as the hysteresis width and the critical threshold or tipping size, have significant impacts on the scale of control intervention --- hysteresis width and critical threshold determine the magnitude of control and duration of application, respectively, for restoring the system to the pre-tipping state. Therefore, it is essential to estimate quantitatively, the key tipping element attributes before one can design a control strategy and assess its impacts.

Small icecap instability is also present in a seasonally varying EBM \cite{north1981energy}, where it results in a sudden jump to an ice-free scenario once the ice extent crosses the bifurcation threshold \cite{north1990multiple}. SICI in the seasonal EBM can manifest as summer sea-ice loss (see \cite{tietsche2011recovery,armour2011reversibility, abbot2011bifurcations} for more details) or winter sea-ice loss \cite{bathiany2016potential,eisenman2009nonlinear,hezel2014modeled}. The optimal control strategy developed here can be generalized to a seasonal EBM with a sea-ice tipping element (refer to the SI for a detailed discussion and example). 

The EBM used in our study is highly idealized in terms of process-level representation; it lacks, for example, a representation of the actual sea-ice thermodynamics and dynamics. Further investigation is necessary to understand how the results obtained here may apply to more physical models with processes such as water vapor, lapse rate, and cloud feedbacks as well as sea-ice thermodynamics. \add{Several extensions of the EBM exist that explicitly account for these additional feedback processes which are diagnosed from global climate models (GCM) \cite{wagner2015climate, beer2022revisiting}, and the current optimal control strategy may be adapted to these more realistic models. Additionally, it may be viable to develop adjoint models for GCMs, which are commonly employed for parameter optimization \cite{lyu2018adjoint, gaikwad2024mitgcm}, in order to estimate the sensitivity of the climate state to the control input. These sensitivities can then be used to design the optimal control based on an appropriately designed cost function as in (\ref{eqn:ebm_opt_1}). However, the adjoint models are often developed based on linearity assumptions and therefore may have limited applicability once the climate system undergoes tipping. Therefore, separate adjoint models would need to be developed for the different regimes of system behavior -- e.g., the ice-free and finite-ice regimes for the Polar sea-ice loss tipping point.} 

More generally, the current approach can be applied to several other tipping elements such as AMOC and West Antarctic Ice Sheet \cite{ditlevsen2023warning}, which also result from saddle-node bifurcations. Saddle-node bifurcations are associated with a universal `normal' form \cite{wigginsintroduction, bury2021deep} that governs the dynamics, and thus physical systems possessing a saddle-node bifurcation will all produce qualitatively similar behavior -- even when the physics underlying those systems are different. Therefore, the policy implications of our study, especially regarding the low relative cost of pre-tipping control could be pertinent to other tipping elements of the earth system.

\backmatter

\bmhead{Supplementary information}
This article has supplementary information. 

\bmhead{Acknowledgements}
This research was supported by the U.S. Department of Energy (DOE), Office of Science, Office of Biological and Environmental Research, Regional and Global Model Analysis program area as part of the HiLAT-RASM project. We also gratefully acknowledge the support from the M1199 NERSC project award. The Pacific Northwest National Laboratory (PNNL) is operated for DOE by Battelle Memorial Institute under contract DE-AC05-76RLO1830. We also thank the two anonymous reviewers who provided constructive feedback and helped improve the paper. 

\section*{Declarations}

\subsection*{Funding}
This research was supported by the U.S. Department of Energy (DOE), Office of Science, Office of Biological and Environmental Research, Regional and Global Model Analysis program area as part of the HiLAT-RASM project.

\subsection*{Data availability}
The data generated for the study is available from the corresponding authors upon request.

\subsection*{Code availability}
Code developed for this research is available here:\begin{verbatim}
https://portal.nersc.gov/cfs/m1199/pkooloth/EBM_control/  \end{verbatim}

\subsection*{Competing interests}
We declare that none of the authors have competing financial or non-financial interests as defined by Nature Portfolio.

\subsection*{Author Contribution}
J.L conceived the research; P.K designed the optimal control and prepared the plots; C.B, J.L, D.D, A.R contributed to the interpretation of results; P.K, and C.B performed numerical studies; P.K, J.L, C.B, D.D and A.R wrote the paper.

\subsection*{Ethics approval and consent to participate}
Not applicable

\subsection*{Consent for publication}
Not applicable


\begin{appendices}

\section{Energy Balance Model}
\label{sec:A}
In the EBM used in our study, the zonally-averaged annual mean surface temperature $T$ evolves as,
\begin{align*}
    C \frac{\partial T}{\partial t} = &(1-\alpha(T))Q_{SW}(x) - (aT + b_0) + F + D\frac{\partial}{\partial x} \left( (1-x^2) \frac{\partial T}{\partial x} \right), 
\end{align*}
subject to the boundary conditions on heat flux at the pole and equator, $$(1-x^2)\frac{\partial T}{\partial x} \big |_{x=0,1} = 0,$$
where $x=\sin{\theta}$
is the latitudinal coordinate. The latitude-dependent incoming annual mean shortwave (SW) radiation is $Q_{SW} = Q_0/4 (1+ 0.482/2(3x^2-1))$ \cite{north1975analytical} and the linearized outgoing longwave radiation (OLR) term is $aT + b_0$. The temperature dependence of the ice-albedo, $\alpha(T)$ is given by

\begin{align*}
    \alpha (T) = 0.62 - \frac{0.3}{1+\exp(-10(T+10))},
\end{align*}
which is a logistic function with $\alpha = 0.32$ for $T \gg -10^\circ$ C and $\alpha = 0.62$ for $T \ll -10^\circ$ C.

\begin{table}[h!]
\centering
\caption{Parameters used in 1D Diffusive Energy Balance Model}
\begin{tabular}{ccc}
Parameter & Description & Value \\
\midrule
$C$ & Ocean mixed-layer heat capacity (W yr/m$^2$/K)  & 8.2  \\
$a$ & OLR Temperature dependence (W/m$^2$/K) & 2.1  \\
$b_0$ & OLR at $T= 0^0$ C (W/m$^2$)& 205\\
$Q_0$ & Solar constant (W/m$^2$)& 1332\\
$D$ & Diffusion coefficient (W/m$^2$/K)& 0.6\\
$F$ & Greenhouse gas forcing (W/m$^2$)& Varies\\
\bottomrule
\label{tab:EBM_params}
\end{tabular}
\end{table}

\section{Variational approach for optimizing control}
\label{sec:B}
The cost function $\mathcal{C}$ is chosen as,
\begin{align}
    \mathcal{C}(T(x,t),u(x,t)) =  \left( w_u \frac{u^2}{\bar{u}^2} + w_T \frac{(T - T_f)^2}{\bar{T}^2} \right), \nonumber \end{align}
where $T_f(x)$ is the ideal temperature profile to be attained via control. The deviation from the target profile and the control are normalized by $\bar{T} = 15^{\circ}$C and $\bar{u} = 0.3$, which represent approximately the mean global temperature and the maximum external albedo control that can be applied, respectively. Note that any variation in the normalization factors can be absorbed into the weight ratio $w_T/w_u$ and therefore our results are not sensitive to the choice of these normalization constants. The optimization problem is then given by,

\begin{align*}
\min_{u(x,t)} \mathcal{L}[u, \; &T] = \min_{u(x,t)}  \int_0^{t_f} \int_0^1 dx \; dt \; \mathcal{C}(T(x,t),u(x,t)) \\
&= \min_{u(x,t)} \int_0^{t_f} \int_0^1 dx \; dt \left( w_u\frac{u^2}{\bar{u}^2} + w_T\frac{(T - T_f)^2}{\bar{T}^2} \right)
    \end{align*}
where $\mathcal{L}[u, T]$ is the total cost functional and $t_f$ is the time period over which the control is to be applied. 
 
Now, using the implicit function theorem on the EBM with control in (\ref{eqn:EBM_ctrl}), the temperature $T$ can be treated as a function of the applied control $u(x,t)$, $x$  and $t$, as such can be expressed as $T(u(x,t),x,t)$. Using variational calculus, the optimal $u(x,t)$ that minimizes $\mathcal{L}[u,T]$ can be computed by setting the first variation in $\mathcal{L}$ to be zero. We describe this here using an arbitrary function $\eta(x,t)$ which goes to $0$ at the domain boundaries and a small constant $\epsilon$. We have,

\begin{align*}
\delta \mathcal{L} [u, T] &= \mathcal{L}[u +  \epsilon \eta, T (u + \epsilon \eta, x, t)] - \mathcal{L}[u, T], \nonumber \\
& = \int_0^{t_f} \int_0^1 dx \; dt \; \delta C,
\end{align*}
where,
\begin{align}
    \delta C = \mathcal{C}(T(u+\epsilon \eta,x,t), u(x,t) + \epsilon \eta) -  \mathcal{C}(T(u,x,t),u(x,t)). \nonumber
\end{align}
Now, $T(u+\epsilon \eta,x,t)$ can be expanded using a Taylor series around $\epsilon=0$ as follows,
\begin{align}
    T(u+\epsilon \eta,x,t) = T(u,x,t) + \frac{\partial T} {\partial u} \epsilon \eta + O(\epsilon^2), \nonumber
\end{align}
and can be substituted into $\delta C$, to obtain,
\begin{align}
 \delta \mathcal{L} = \int_0^{t_f} \int_0^1 dx \; dt \; 2 \left(  w_u \frac{u}{\bar{u}^2} +  w_T \frac{T - T_f}{\bar{T}^2 } \frac{\partial T} {\partial u}\right) \epsilon \eta + O(\epsilon^2). \nonumber
    \end{align}
 Now, for the first variation to vanish we need,

\begin{align*}
\lim_{\epsilon \to 0} \frac{1}{\epsilon} \delta \mathcal{L} = &\lim_{\epsilon \to 0} \int_0^{t_f} \int_0^1 dx \; dt \; 2 \left( w_u \frac{u}{\bar{u}^2} +  w_T \frac{T - T_f}{\bar{T}^2 } \frac{\partial T} {\partial u}\right) \eta + O(\epsilon) \nonumber \\
&= \int_0^{t_f} \int_0^1 dx \; dt \; 2 \left( w_u \frac{u}{\bar{u}^2} +  w_T\frac{T - T_f}{\bar{T}^2 } \frac{\partial T} {\partial u}\right) \eta \\
&= 0. 
\end{align*}
And now since $\eta$ is arbitrary, we get an equation for the optimal $u$ given by,
\begin{align*}
    \frac{u}{\bar{u}^2} + \frac{w_T}{w_u}\left( \frac{T - T_f}{\bar{T}^2 } \right) \frac{\partial T} {\partial u} = 0, 
\end{align*}
which is the Euler-Lagrange equation for this minimization problem. So as to compute $u(x,t)$, $T_u(x,t)$ must be known; this can be achieved by solving an additional equation obtained by taking a partial derivative of the EBM with respect to $u$,
\begin{align}
    C\frac{\partial T_u}{\partial t} = (-\alpha'(T)T_u - &1)Q_{SW}(x) -aT_u  + D\frac{\partial}{\partial x} \left( (1-x^2) \frac{\partial T_u}{\partial x} \right)   
\end{align}
subject to boundary conditions, $(1-x^2)\frac{\partial T_u}{\partial x} \big |_{x=0,1} = 0$.

\subsection{Bounds for weight ratio} The bounds for $w_T/w_u$ for a given set of parameters for the EBM are obtained by running the controlled EBM for a range of $w_T/w_u$ (with a step size of 0.01). The ice-free starting state is used as this scenario requires the largest applied control among the three starting states considered. The upper bound ($\lambda_{max}$) is the maximum value of $w_T/w_u$ for which $u(x,t) < 1 - \alpha(x,t)$ for all times $t$. The lower bound ($\lambda_{min}$) is the value of $w_T/w_u$ such that for $w_T/w_u < \lambda_{min}$, the applied control fails to adjust to the target profile.

\section{Numerical simulations}
\label{sec:C}
Numerical simulations were carried out in Dedalus \cite{2020PhRvR...2b3068B}-- a spectral PDE-solver. 

\subsection{Steady solutions}
The steady-state solutions, both stable and unstable, for the EBM were computed by solving the nonlinear boundary value problem using a Legendre basis for $x$ with $256$ basis elements and iterated using the Newton method. 

\subsection{Non-equilibrium solutions}

The time-varying problem with GHG forcing, with or without control, is an initial value problem and is solved using Runge Kutta 443 (RK443) time integration and a Legendre basis in $x$ with 256 basis elements with 3/2 dealiasing.




\end{appendices}



\end{document}